\documentclass[3p]{elsarticle}
\usepackage{epsfig,amsmath,amssymb,graphicx}

\newtheorem{theorem}{Theorem}

\usepackage{setspace}
\usepackage{epstopdf}
\usepackage{xcolor}
\newcommand{\diag}{\mathop{\mathrm{diag}}}
\begin{document}
\begin{frontmatter}
\title{Synchronization of an evolving complex hyper-network}
\author[zyw]{Zhaoyan Wu}
\ead{zhywu7981@gmail.com}
\author[jqd1,jqd2]{Jinqiao Duan}
\ead{jduan@ipam.ucla.edu}
\author[xcf]{Xinchu Fu\corref{cor}}
\ead{xcfu@shu.edu.cn}
\cortext[cor]{Corresponding author.}
\address[zyw]{College of Mathematics and Information Science, Jiangxi Normal University,
Nanchang 330022, China}
\address[jqd1]{Institute for Pure and Applied Mathematics, University of California, Los Angeles, CA 90095-7121, USA}
\address[jqd2]{Department of Applied Mathematics, Illinois Institute of Technology, Chicago, IL 60616, USA}
\address[xcf]{Department of Mathematics, Shanghai University, Shanghai 200444, China}

\begin{abstract}
\noindent In this paper, the synchronization in a hyper-network of coupled dynamical systems is investigated for the first time. An evolving hyper-network model is proposed for better describing some complex systems. A concept of joint degree is introduced, and the evolving mechanism of hyper-network is given with respect to the joint degree. The hyper-degree distribution of the proposed evolving hyper-network is derived based on a rate equation method and obeys a power law distribution. Furthermore, the synchronization in a hyper-network of coupled dynamical systems is investigated for the first time. By calculating the joint degree matrix, several simple yet useful synchronization criteria are obtained and illustrated by several numerical examples.
\end{abstract}
\end{frontmatter}

\section{Introduction}

Complex networks \cite{DJW,ALB,RA,MEJN,QC} have been used to model the real complex systems with a large number of interacting individuals, such as
Internet \cite{AV}, world trade web \cite{AM}, metabolic networks \cite{JH}, coauthor networks \cite{MEJ}, and so on. The individuals in complex systems are denoted by nodes in complex networks, and the interactions between individuals are denoted by edges between nodes. For example, in a collaboration network, the scientists can be regarded as nodes and their joint papers can be treated as edges. From this collaboration network, we can find the collaborative relationship between any two scientists. However, many papers have more than two coauthors. Therefore, the collaborative relationship among any three or more scientists can not be described in the above simple collaboration network. Then, what kind of networks can solve this problem is an interesting issue, and the complex hyper-networks corresponding to hyper-graphs emerge in this context.

In complex networks, an edge connects only two nodes, however, an hyper-edge of complex hyper-networks can contain more than two nodes, which can describe the multifaceted collaborative relationship appropriately. Though a hyper-edge can connect an arbitrary number of nodes, it is often useful to study hyper-networks where each hyper-edge  connects the same number of nodes: a $k$-uniform hyper-complex is a hyper-network in which each hyper-edge   connects exactly $k$ nodes.

There has been a growing interest in the research of hyper-networks [10-14] 
recently. For example, in virtual enterprises, to respond to market changes rapidly and exploit unexpected business opportunities efficiently,
the virtual breeding environment (VBE) actors collaborate, and share competencies, skills as well as resources \cite{CLM}. In order to describe the relationship among VBE actors, a hyper-graph (a hyper-network) is proposed as
a meaningful logical structure, in which a hyper-path (i.e., a hyper-edge) represents the structure underlying a minimal cluster of enterprises. By the hyper-graph model, the formation process of a virtual enterprise can be presented and the pre-identified business opportunity may be caught.

Recently, an evolving hyper-network model is introduced to describe real-life systems with the hyper-network characteristics \cite{WJW}. Two evolving mechanisms with respect to a hyper-degree, namely hyper-edge growth and preferential attachment, are proposed to construct the hyper-network mode. The hyper-degree is defined as the number of the hyper-edge attached to that node. In this paper, a different evolving $k$-uniform hyper-network model is introduced, whose evolving mechanism involves the joint degree of $k-1$ nodes, which is defined as the number of the hyper-edges attached to the $k-1$ nodes.

On the other hand, synchronization as a typical collective dynamical behavior of complex networks has drawn considerable attentions recently [15-27]. However, less attention has been paid to   synchronization of complex hyper-networks. Thus in this paper, synchronization of the 3-uniform hyper-network will be investigated for the first time. The rest of this paper is organized as follows. In Section 2, an evolving hyper-network model is introduced and several topological characteristics are studied. In Section 3, synchronization of a 3-uniform hyper-network is investigated and several synchronization criteria are obtained. In Section 4, several numerical examples are provided to verify the effectiveness of derived results. Conclusions are drawn in Section 5.

\section{An evolving hyper-network model}

An evolving mechanism with respect to joint degree to construct the $k$-uniform hyper-network model is proposed in this section. In the scale-free network model put forward by Barab\'{a}si and Albert \cite{ALB}, two simple evolving mechanisms, i.e., growth and preferential attachment, are proposed to construct the network model. Inspired by \cite{ALB}, the following generation algorithm are proposed to construct the $k$-uniform hyper-network model:

\noindent(1) Growth: the hyper-network starting with $m_0~(m_0\geq k)$ nodes, 
at every time step, we add one node and $m$ hyper-edges to the existing hyper-network, where $m\leq C_{m_0}^{k-1}=\frac{m_0!}{(k-1)!(m_0-k+1)!}$.

\noindent(2) Preferential attachment: the probability $\Pi_{i_1i_2...i_{k-1}}$of a new hyper-edge contains the new node and selected $k-1$ nodes in the existing hyper-network depends on the joint degree of the $k-1$ nodes at time $t$. Let $d_{i_1i_2...i_{k-1}}$ denotes the joint degree of nodes $i_1,i_2,\cdots,i_{k-1}$, which is defined as the number of hyper-edges containing nodes $i_1,i_2,\cdots,i_{k-1}$, then
\begin{equation*}
\Pi_{i_1i_2...i_{k-1}}=\frac{d_{i_1i_2...i_{k-1}}}{\sum\limits_{1\leq l_1<l_2<...<l_{k-1}\leq m_0+t-1}d_{l_1l_2...l_{k-1}}},
\end{equation*}
where $1\leq i_1< i_2< \cdots < i_{k-1}\leq m_0+t-1$.

\begin{figure}
\centerline{\includegraphics[height=60mm,width=80mm]{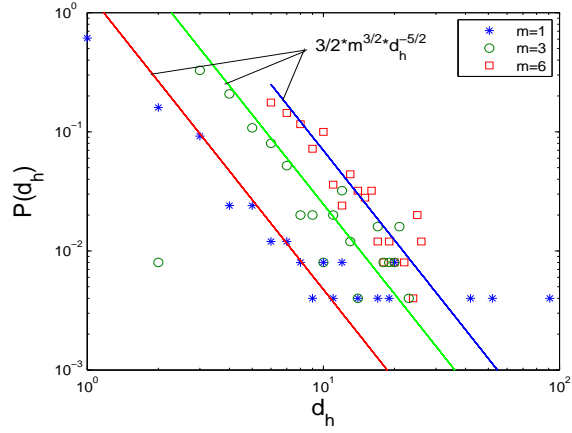}}
\caption{Hyper-degree distribution $p(d_h)$ versus node
hyper-degree $d_h$ on a logarithmic scale with the different parameters. The node number is $N=250$.}\label{eps2}
\end{figure}

Then, after $t$ time steps, this constructs a $k$-uniform hyper-network with $m_0+t$ nodes and $mt$ hyper-edges. Define the hyper-degree $d_h^i$ of node $i$ as the number of the hyper-edges containing node $i$ and node-node distance as the number of the hyper-edges in the shortest paths between two nodes. According to the generation algorithm and the relation between hyper-degree and joint degree, when a new node enter into the hyper-network, the probability of node $i$ with hyper-degree $d_h^i$ acquires a hyper-edge is

\begin{align*}
\Pi(d_h^i)=&\sum_{1\leq i_1<\cdots<i<\cdots<i_{k-1}\leq m_0+t-1}\Pi_{i_1\cdots i\cdots i_{k-1}}\\
=&\frac{(k-1)d_h^i}{mkt}.
\end{align*}
Let $i$ be the new node added to the hyper-network at time $t_i$, and $p(d_h,t_i,t)$ be
the probability that node $i$ has hyper-degree $d_h^i$ when it is being picked up at time $t~(t\geq t_i)$. Suppose that $d_h^i$ is a continuous variable, then probability $\Pi(d_h^i)$ can be viewed as a continuous rate of change of $d_h^i$, i.e.,
\begin{equation}\label{eq5}
\frac{\partial d_h^i}{\partial t}=a\Pi(d_h^i)=\frac{a(k-1)d_h^i}{mkt}
\end{equation}
where $a$ is a constant and denotes the rate of change of $d_h^i$. Since the new node $i$ brings in $m$ new hyper-edges, which has hyper-degree $m$ at
time $t_i$, so the change of hyper-degree at $t_i$ is $a=m$. Solving equation (\ref{eq5}) with initial condition $d_h^i(t_i)=m$, one has
$$
d_h^i(t)=m(\frac{t}{t_i})^{\frac{k-1}{k}}
$$
and
$$
t_i=(\frac{m}{d_h^i})^{\frac{k}{k-1}}t.
$$
Assuming that the time variables $t_i$ have a uniform distribution with $P(t_i)=1/t$, i.e., the new nodes are being added at equal time intervals, then one has
\begin{align*}
P(d_h^i(t)<d_h)=&P(t_i>(\frac{m}{d_h^i})^{\frac{k}{k-1}}t)\\
=&1-P(t_i<(\frac{m}{d_h^i})^{\frac{k}{k-1}}t)\\
=&1-(\frac{m}{d_h^i})^{\frac{k}{k-1}}.
\end{align*}
Therefore, hyper-degree distribution $P(d_h^i(t)=d_h)$ can be derived as
\begin{equation*}
P(d_h)=\frac{\partial P(d_h^i(t)<d_h)}{\partial d_h}=\frac{k}{k-1}m^{\frac{k}{k-1}}d_h^{-2-\frac{1}{k-1}},
\end{equation*}
which is the same as that of BA scale-free network when $k=2$. Fig. \ref{eps2} shows the hyper-degree distribution of the 3-uniform hyper-network models with $m_0=4$, $m=1,3,6$ and $t=196$,
from which one can find that the 3-uniform hyper-network has power distribution property, i.e., scale-free property. Table I shows the average path length of the 3-uniform hyper-network models with $m_0=4$ and different pair of $t$ and $m$, which exhibits small world property.

\begin{table}
\center{\begin{tabular}{lcccccc}
 & \vline & $N=50$ & $N=100$ & $N=150$ & $N=200$ & $N=250$  \\
\hline
$m=1$ & \vline & 2.6176 & 2.8893 & 2.9459 & 3.1752 & 3.2785 \\
$m=2$ & \vline & 2.0539 & 2.2786 & 2.3688 & 2.4823 & 2.5124 \\
$m=4$ & \vline & 1.7683 & 1.9461 & 2.0135 & 2.0849 & 2.1437 \\
\hline
\end{tabular}}
\caption{The average path length of the 3-uniform hyper-network with $m_0=4$ and different pair of $N$ and $m$, where $N$ is the node number, i.e., $N=t+m_0$.}
\end{table}

\section{Synchronization in hyper-network of coupled dynamical systems}
Complex network consisting of $N$ identical linearly and diffusively coupled nodes can be described by
\begin{equation}\label{eq1}
\dot{x}_i(t)=f(x_i(t))+\varepsilon\sum_{j=1,j\neq i}^Na_{ij}\Gamma(x_j(t)-x_i(t)),~i=1,2,\cdots,N,
\end{equation}
where $x_i(t)=(x_{i1}(t),x_{i2}(t),\cdots,x_{in}(t))^T\in R^n$ is the state variable of node $i$, $f:R^n\rightarrow R^n$ represents the local dynamics of an isolated node, $\varepsilon>0$ is coupling strength, $\Gamma\in R^{n\times n}$ is inner coupling matrix. Matrix $A=(a_{ij})_{N\times N}$ is outer coupling matrix, which denotes the network topology and is defined as, if there is a connection from node $i$ to node $j~(i\neq j)$, then $a_{ij}\neq0$, otherwise, $a_{ij}=0$. According to the definition of the coupling matrix $A$, the network (\ref{eq1}) covers a variety of models ranging from simple weightless undirected networks to more complicated weighted directed networks.

Similar to the network (\ref{eq1}), a $k$-uniform hyper-network of $N$ coupled systems can be described by
\begin{equation}\label{eq2}
\dot{x}_{i_1}(t)=f(x_{i_1}(t))+\varepsilon\sum_{i_2,\cdots,i_k=1}^Na_{i_1i_2\cdots i_k}\sum_{j=2}^{k}\Gamma(x_{i_j}(t)-x_{i_1}(t)),~i_1=1,2,\cdots,N,
\end{equation}
where $i_l~(l=1,2,\cdots,k)$ are different each other and $a_{i_1i_2\cdots i_k}$ is defined as: if there exists a hyper-edge contains nodes $i_1,i_2,\cdots,i_k$, then $a_{i_1i_2\cdots i_k}=1$, otherwise, $a_{i_1i_2\cdots i_k}=0$. For $k=3$, the hyper-network (\ref{eq2}) can be simplified as
\begin{equation}\label{eq3}
\dot{x}_i(t)=f(x_i(t))+\varepsilon\sum_{j=1}^N\sum_{k=1}^Na_{ijk}\Gamma(x_j(t)+x_k(t)-2x_i(t)).
\end{equation}

In the following, consider synchronization of the 3-uniform hyper-network (\ref{eq3}), which can be rewritten as
\begin{equation}\label{eq51}
\dot{x}_i(t)=f(x_i(t))+\varepsilon\sum\limits_{j=1}^N(\sum\limits_{k=1}^Na_{ijk})\Gamma(x_j(t)-x_i(t))+\varepsilon\sum\limits_{k=1}^N(\sum\limits_{j=1}^Na_{ijk})\Gamma(x_k(t)-x_i(t)).
\end{equation}
Let $d_{ij}$ be the joint degree of nodes $i$ and $j~(j\neq i)$, which is defined as the number of hyper-edges containing nodes $i$ and $j$. Then one has
\begin{equation*}
  d_{ij}=\sum\limits_{k=1}^Na_{ijk}=\sum\limits_{k=1}^Na_{ikj}=\cdots=\sum\limits_{k=1}^Na_{kji},
\end{equation*}
and rewrite (\ref{eq51}) as
\begin{equation*}
\dot{x}_i(t)=f(x_i(t))+\varepsilon\sum\limits_{j=1,j\neq i}^Nd_{ij}\Gamma(x_j(t)-x_i(t))+\varepsilon\sum\limits_{k=1,k\neq i}^Nd_{ik}\Gamma(x_k(t)-x_i(t)).
\end{equation*}
Thus,
\begin{equation}\label{eq7}
\dot{x}_i(t)=f(x_i(t))+2\varepsilon\sum\limits_{j=1,j\neq i}^Nd_{ij}\Gamma(x_j(t)-x_i(t)).
\end{equation}
 Define the diagonal elements of $D$ as follows
$$
d_{ii}=-\sum_{j=1,j\neq i}^Nd_{ij}=-\sum_{j=1,j\neq i}^Nd_{ji}.
$$
Then, equation (\ref{eq7}) gives
\begin{equation}\label{eq8}
\dot{x}_i(t)=f(x_i(t))+2\varepsilon\sum\limits_{j=1}^Nd_{ij}\Gamma x_j(t).
\end{equation}
In the subsequent studies, the hyper-network considered is always assumed to be connected, i.e., any pair of nodes is reachable along hyper-edge. It is easy to verify that the joint degree matrix $D=(d_{ij})$ is irreducible and its eigenvalues are $0=\lambda_1>\lambda_2\geq\cdots\geq \lambda_N$. Our objective here is to synchronize the network (\ref{eq8}) with a given orbit $s(t)$, i.e.,
\begin{equation}\label{eq9}
  x_1(t)=x_2(t)=\cdots=x_N(t)=s(t),~~~~as~t\rightarrow\infty,
\end{equation}
where $s(t)$ is a solution of an isolated node, namely, $\dot{s}(t)=f(s(t))$. Here, $s(t)$ can be an equilibrium point, a periodic orbit, or even a chaotic attractor.

Let $x_i(t)=s(t)+\eta_i(t)$ and linearize (\ref{eq8}) about $s(t)$.  This leads to
\begin{equation}\label{eq10}
  \dot{\eta}_i(t)=Df(s(t))\eta_i(t)+2\varepsilon\sum\limits_{j=1}^Nd_{ij}\Gamma\eta_j(t),
\end{equation}
where $Df(s(t))$ is is the Jacobian of $f$ on $s(t)$. Then refer to the proofs of lemmas 1 and 2 in \cite{WXF1}, one has the following theorems.

\begin{theorem}
Let $0=\lambda_1>\lambda_2\geq\cdots\geq \lambda_N$ be the eigenvalues of the joint degree matrix $D=(d_{ij})$. If the following $N-1$ $n$-dimensional linear time-varying systems are exponentially stable
  \begin{equation}\label{eq102}
    \dot{\xi}(t)=(Df(s(t))+2\varepsilon\lambda_k\Gamma)\xi(t),~k=2,3,\cdots,N,
  \end{equation}
then the synchronized states (\ref{eq9}) are exponentially stable.
\end{theorem}
\textbf{Proof.} The proof will be given in the Appendix. $\Box$

\begin{theorem}
Suppose that there exists an $n\times n$ diagonal matrix $P>0$ and two constants $\bar{\delta}<0$ and $\tau>0$ such that
\begin{equation}\label{eq101}
  (Df(s(t))+\bar{\delta}\Gamma)^TP+P(Df(s(t))+\bar{\delta}\Gamma)\leq -\tau I_n,
\end{equation}
for all $\delta\leq \bar{\delta}$, where $I_n$ is an $n\times n$ identity matrix. If
\begin{equation}\label{eq11}
  2\varepsilon\lambda_2\leq \bar{\delta},
\end{equation}
then the synchronized states (\ref{eq9}) are exponentially stable.
\end{theorem}
\textbf{Proof.} The proof will be given in the Appendix. $\Box$

It is clear that the inequality (\ref{eq11}) is equivalent to
\begin{equation}\label{eq12}
  \varepsilon\geq \frac{\bar{\delta}}{2\lambda_2},
\end{equation}
Therefore, synchronizability of the 3-uniform hyper-network (\ref{eq51}) with respect
to a given coupling matrix can be characterized by the second-largest eigenvalue $\lambda_2$ of the joint degree matrix.

\noindent \textbf{Remark 1.} Similar to the above discussions, synchronization of general $k$-uniform hyper-network can be investigated as well. In fact, refer to the definition of joint degree, hyper-network (\ref{eq2}) can be simplified as
\begin{equation*}
\dot{x}_i(t)=f(x_i(t))+(k-1)\varepsilon\sum\limits_{j=1}^Nd_{ij}\Gamma x_j(t).
\end{equation*}

\section{Numerical simulations}
Firstly, consider a hyper-network consisting of 100 coupled Chua's oscillator \cite{CLO}, which can be described by
\begin{equation*}
\left(%
  \begin{array}{c}
  \dot{s}_1 \\
  \dot{s}_2 \\
  \dot{s}_3
  \end{array}
  \right)=
  \left(%
  \begin{array}{c}
  \alpha (s_2-s_1-f(s_1)) \\
  s_1-s_2+s_3 \\
  -\beta s_2-\gamma s_3
  \end{array}
  \right)
\end{equation*}
where $f(s_1)=bs_1+0.5(a-b)(|s_1+1|-|s_1-1|)$ is a piecewise linear function, $\alpha>0$, $\beta>0$, $\gamma>0$, $a<0$ and $b<0$.

\begin{figure}
\centerline{\includegraphics[height=50mm,width=100mm]{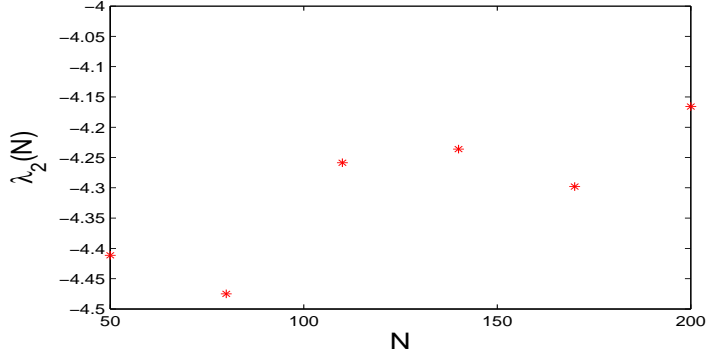}}
\caption{The second-largest eigenvalue of the joint degree matrix with $m=3$.}\label{eps3}
\end{figure}

\begin{figure}
\centerline{\includegraphics[height=60mm,width=100mm]{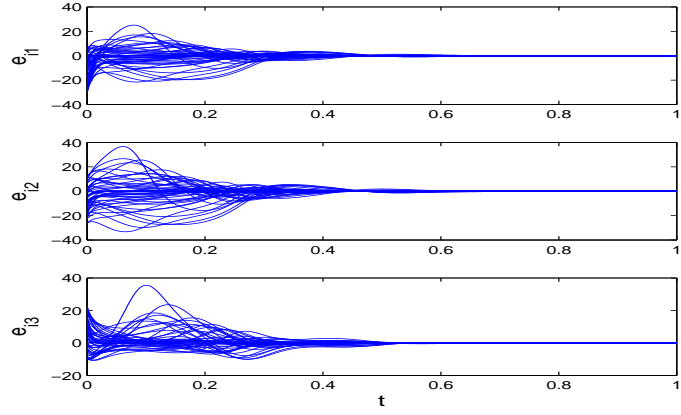}}
\caption{The synchronization errors.}\label{eps4}
\end{figure}

Fig. \ref{eps3} shows the second-largest eigenvalue of the joint degree matrix $D$ generating from the above evolving algorithm with $m_0=4$ and $N=100$, where $\lambda_2(N)$ is obtained by
averaging the results of 10 runs. Then one has $\lambda_2<-4$. In the numerical simulation, choose $\alpha=10$, $\beta=15$, $\gamma=0.0385$, $a=-1.27$ and $b=-0.68$. Refer to the discussion in \cite{WXF1}, one can choose $\bar{\delta}=a$ such that the inequality (\ref{eq101}) holds. Then one can choose $\varepsilon=0.5$ such that the inequality (\ref{eq12}) holds. That is to say, the hyper-network generating from the evolving algorithm with $m=3$ can achieve synchronization with any initial values. Fig. \ref{eps4} shows the synchronization errors.
\begin{figure}
\centerline{\includegraphics[height=60mm,width=100mm]{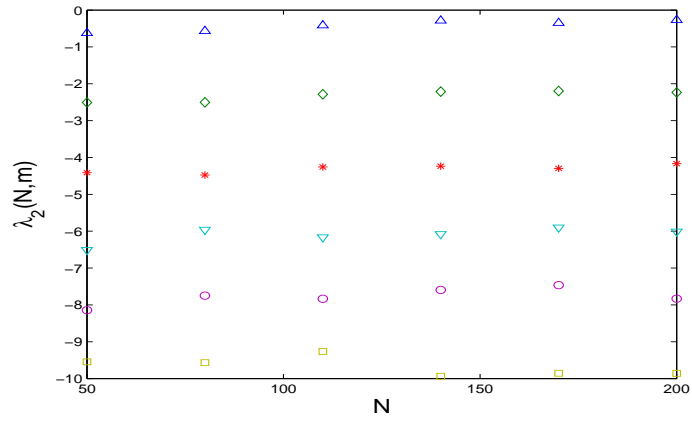}}
\caption{The second-largest eigenvalue of the joint degree matrix. The `$\triangle$' denotes $m=1$, `$\diamond$'denotes $m=2$, `$\ast$' denotes $m=3$, `$\nabla$' denotes $m=4$, `$\circ$' denotes $m=5$ and `$\square$' denotes $m=6$.}\label{eps5}
\end{figure}

\begin{figure}
\centerline{\includegraphics[height=40mm,width=55mm]{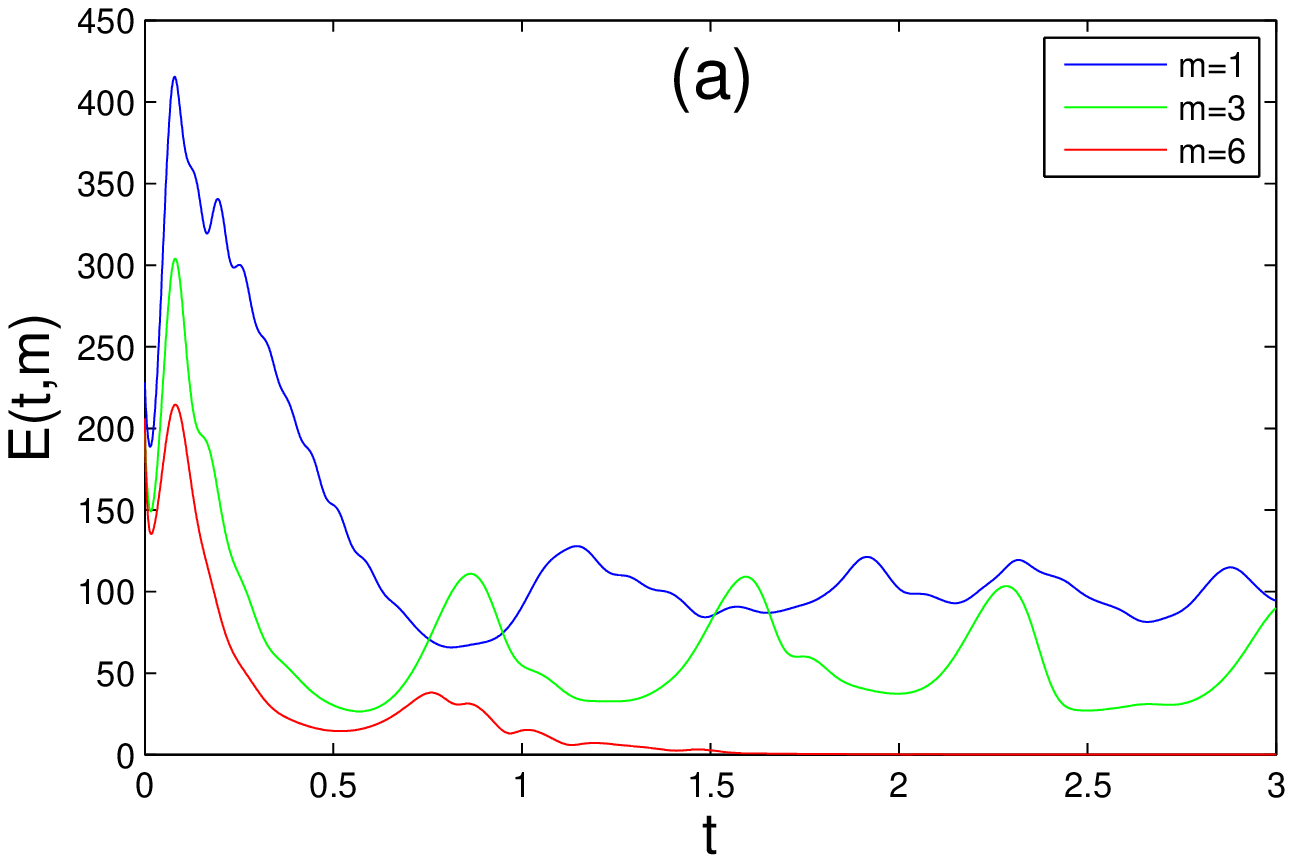}
\includegraphics[height=40mm,width=55mm]{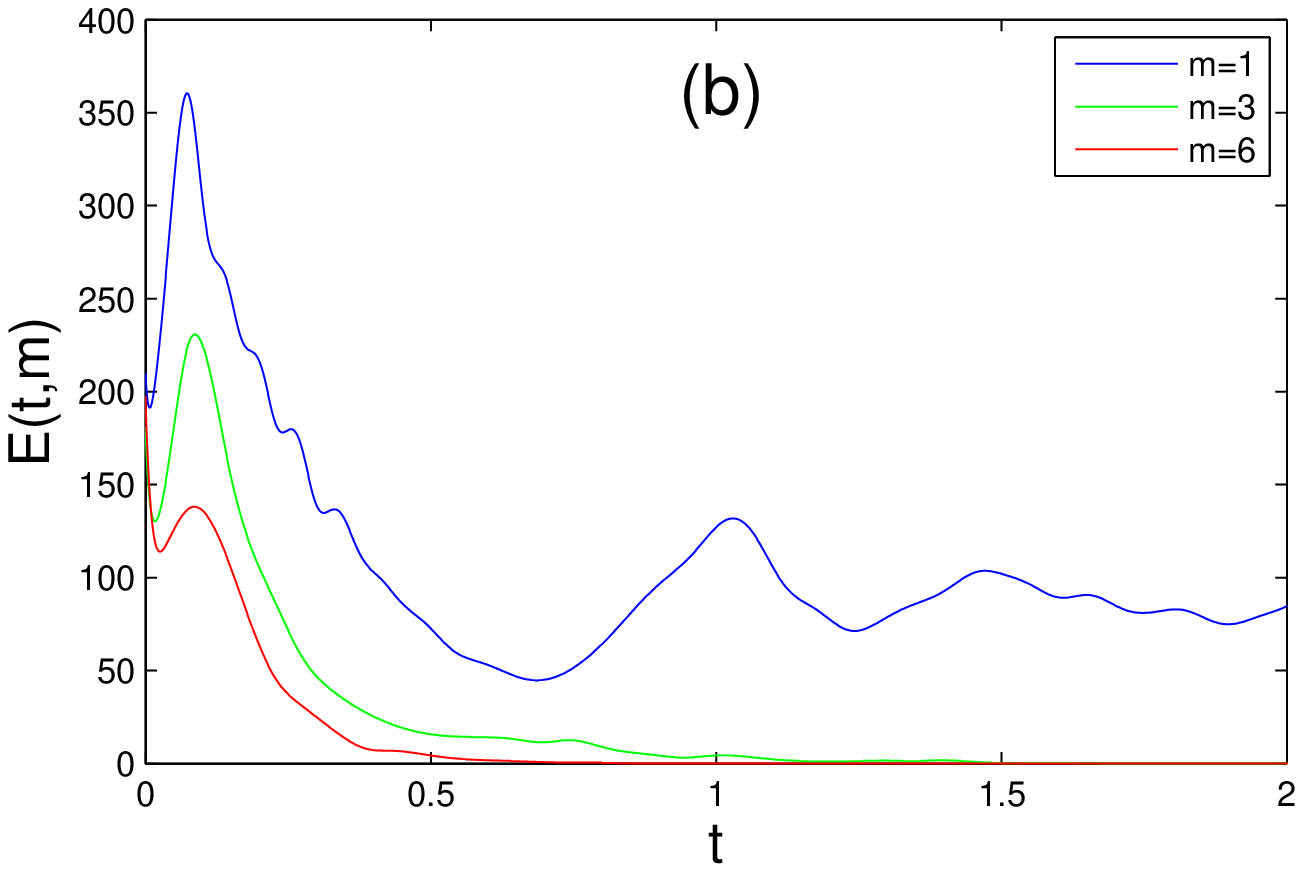}
\includegraphics[height=40mm,width=55mm]{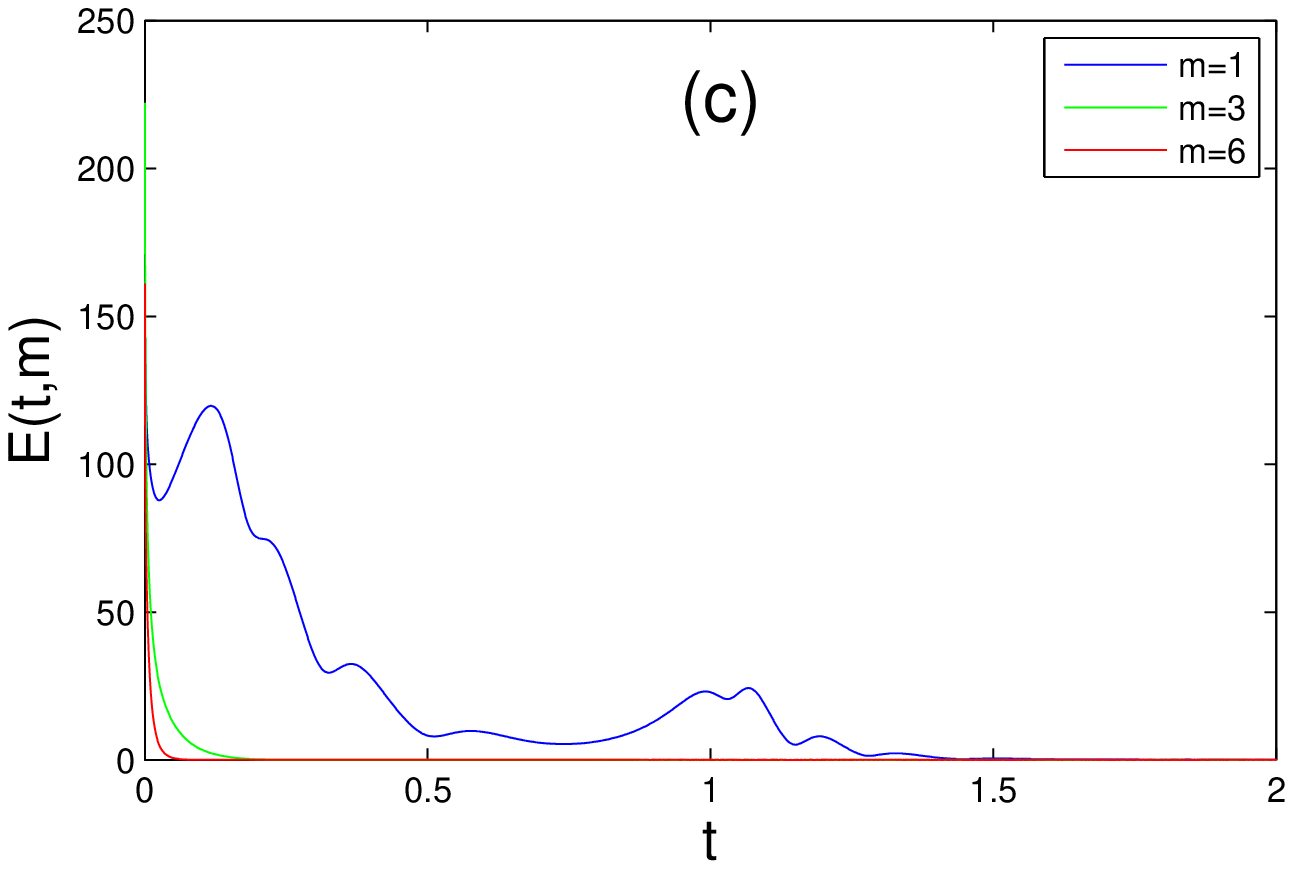}}
\caption{Synchronization errors $E=\sqrt{\sum\limits_{i=2}^N(x_i(t)-x_1(t))^T(x_i(t)-x_1(t))}$. Case (a): $\varepsilon=0.3$, (b): $\varepsilon=0.5$, (c): $\varepsilon=5$.}\label{eps6}
\end{figure}

Secondly, consider the synchronizability of 3-uniform hyper-network with different pair of $m$ and $N$. Fig. \ref{eps5} shows the second-largest eigenvalues of the joint degree matrix with different pair of $N$ and $m$, where $\lambda_2(N,m)$ is obtained by averaging the results of 10 runs. It is clear that $\lambda_2(N,1)>\lambda_2(N,2)>\cdots>\lambda_2(N,6)$, which means the hyper-network with larger $m$ has stronger synchronizability. Fig. \ref{eps6} shows the synchronization errors with $N=100$ and different coupling strength $\varepsilon=0.3$, $\varepsilon=0.5$ and $\varepsilon=5$, from which one can easily find that the hyper-network with $m=6$ has stronger synchronizability.

\section{Conclusions}

An evolving mechanism with respect to joint degree is proposed to construct a $k$-uniform hyper-network model. Based on a rate equation method, the hyper-degree distribution of the hyper-network is derived, which obeys a power law distribution. Moreover, a complex   hyper-network coupled with dynamical systems is introduced. By defining the joint degree of two nodes, the hyper-network can be simplified and its synchronization is investigated for the first time. Further, synchronizability of 3-uniform hyper-network with different pair of $m$ and $N$ is considered by calculating the second-largest eigenvalue $\lambda_2$ of the joint degree matrix.

\section*{Acknowledgement}
This research was jointly supported by the NSFC grant 11072136, the Shanghai Univ. Leading Academic Discipline Project
(A.13-0101-12-004) and Natural Science Foundation of Jiangxi Province of China (20122BAB211006). The authors would also
like to thank the anonymous referees for their helpful comments and suggestions.

\section*{Appendix}

\noindent \emph{Proof of Theorem 1:} Let $\eta(t)=(\eta_1(t),\eta_2(t),\cdots,\eta_N(t))^T\in R^{N\times n}$, then equation (\ref{eq10}) can be written as
\begin{align*}
  \dot{\eta}(t)=\eta(t)(Df(s(t)))^T+2\varepsilon D\eta(t)\Gamma^T.
\end{align*}
Choose an unitary matrix $\Phi=(\phi_1,\phi_2,\cdots,\phi_N)$ such that
\begin{align*}
  A\phi_k=\lambda_k\phi_k,~k=1,2,\cdots,N.
\end{align*}
Let $\Lambda=\diag(\lambda_1,\lambda_2,\cdots,\lambda_N)$ and $\eta(t)=\Phi v(t)$ with $v(t)=(v_1(t),v_2(t),\cdots,v_N(t))^T\in R^{N\times n}$, then one has
\begin{align*}
  \dot{v}(t)=v(t)(Df(s(t)))^T+2\varepsilon \Lambda v(t)\Gamma^T
\end{align*}
and
\begin{align*}
  \dot{v}_k(t)=(Df(s(t))+2\varepsilon\lambda_k\Gamma) v_k(t),~k=1,2,\cdots,N.
\end{align*}
Refer to the proof of Lemma 1 in Ref. \cite{WXF1}, $\lambda_1=0$ corresponds to the synchronization of the system states. Then if the $N-1$ $n$-dimensional linear time-varying systems (\ref{eq102}) are exponentially stable, the synchronized states (\ref{eq9}) is exponentially stable.
\vskip 10pt
\noindent \emph{Proof of Theorem 2:} Consider the following Lyapunov functions
\begin{align*}
  V_k(t)=\xi^T(t)P\xi(t),~k=2,3,\cdots,N.
\end{align*}
Calculate the derivative of $V(t)$ along (\ref{eq102}) gives
\begin{align*}
  \dot{V}_k(t)=\xi^T(t)((Df(s(t))+2\varepsilon\lambda_k\Gamma)^TP+P(Df(s(t))+2\varepsilon\lambda_k\Gamma))\xi(t).
\end{align*}
From inequalities (\ref{eq101}) and (\ref{eq11}), one has
\begin{align*}
  \dot{V}_k(t)\leq&\xi^T(t)((Df(s(t))+\bar{\delta}\Gamma)^TP+P(Df(s(t))+\bar{\delta}\Gamma))\xi(t)\\
  \leq&-\tau \xi^T(t)\xi(t)\leq-\frac{\tau}{\lambda_P}V_k(t),
\end{align*}
where $\lambda_P$ is the largest eigenvalue of $P$, which implies
\begin{align*}
  V_k(t)\leq \exp\{-\frac{\tau}{\lambda_P}\}V(0),~k=2,3,\cdots,N.
\end{align*}
That is, the $N-1$ time-varying systems (\ref{eq102}) are exponentially stable, i.e., the synchronized states (\ref{eq9}) is exponentially stable.

\end{document}